\documentclass[preprintnumbers,nofootinbib,superscriptaddress,showpacs,aps,prd]{revtex4}

\usepackage{amsmath,amssymb,bm}
\usepackage{graphicx}

\newcommand{\ff}{f\hspace{-0.4em}f}
\newcommand{\msbar}{\overline{\text{MS}}}
\newcommand{\mbar}{\overline{m}}
\newcommand{\mubar}{\bar{\mu}}

\begin{document}

\preprint{MZ-TH/11-11}
\preprint{ZU-TH~10/11}

\title{Precision predictions for the $\bm{t\bar{t}}$ production cross section at hadron 
  colliders}

\author{Valentin Ahrens}
\affiliation{Institut f\"ur Physik (THEP), Johannes Gutenberg-Universit\"at, D-55099
  Mainz, Germany}
\author{Andrea Ferroglia}
\affiliation{New York City College of Technology, 300 Jay Street, Brooklyn, NY 11201, USA}
\author{Matthias Neubert}
\author{Ben D.~Pecjak}
\affiliation{Institut f\"ur Physik (THEP), Johannes Gutenberg-Universit\"at, D-55099
  Mainz, Germany}
\author{Li Lin Yang}
\affiliation{Institute for Theoretical Physics, University of Z\"urich, CH-8057 Z\"urich,
  Switzerland}

\date{\today}

\begin{abstract}
  \noindent
  We make use of recent results in effective theory and higher-order
  perturbative calculations to improve the theoretical predictions of
  the top-quark pair production cross section at hadron colliders. In
  particular, we supplement the fixed-order NLO calculation with
  higher-order corrections from soft gluon resummation at NNLL accuracy.
  Uncertainties due to power corrections to the soft limit are
  estimated by combining results from single-particle inclusive and pair 
  invariant-mass kinematics.  We present our predictions as functions of 
  the top-quark mass in both the pole scheme and the $\msbar$ scheme. We 
  also discuss the merits of using threshold masses as an alternative, and 
  calculate the cross section with the top-quark mass defined in the 1S 
  scheme as an illustrative example. 
\end{abstract}


\maketitle

\section{Introduction} 

The total $t\bar t$ production cross section is an important observable at hadron
colliders such as the Tevatron and LHC. For instance, it provides information about the
top-quark mass, which is an input for electroweak fits \cite{Flacher:2008zq} used to
constrain the mass of the Higgs boson. Extractions of the top-quark mass from the
production cross section have the advantage that the perturbative calculations used in the
analysis are carried out in a well-defined renormalization scheme for the top-quark mass.
The pole mass as well as the $\msbar$ mass have already been extracted from the production
cross section at the Tevatron \cite{Abazov:2011pt}. The use of a short-distance
mass such as the $\msbar$ mass is theoretically favored over the pole mass, which can be
defined only up to a renormalon ambiguity of order $\Lambda_{\rm QCD}$. Moreover, it has
been proposed in \cite{Langenfeld:2009wd} that the apparent convergence and scale
uncertainties of the perturbative series for the total cross section is improved in the
$\msbar$ scheme compared to the pole scheme, already at low orders in the perturbative
expansion.

According to the QCD factorization theorem \cite{Collins:1989gx}, the hadronic cross
section can be obtained from the partonic one after convolution with parton distribution
functions (PDFs). For $t\bar{t}$ production, it is conventional to express this
factorization theorem in the form
\begin{align}
\label{eq:ff}
  \sigma(s,m_t) = \frac{\alpha_s^2(\mu_f)}{m_t^2} \sum_{i,j} \int_{4m_t^2}^s \frac{d\hat{s}}{s}
  \, \ff_{ij} \bigg( \frac{\hat{s}}{s}, \mu_f \bigg) \, f_{ij} \bigg( \frac{4
    m_t^2}{\hat{s}}, \mu_f \bigg) \, ,
\end{align}
where $\mu_f$ is the factorization scale, $i,j \in \{q,\,\bar q,\,g\}$, and the parton 
luminosities are defined as
\begin{align}
  \ff_{ij}(y,\mu_f) = \int_y^1 \frac{dx}{x} \, f_{i/N_1}(x,\mu_f) \, f_{j/N_2}(y/x,\mu_f)
  \, .
\end{align}
Here and throughout this letter the running coupling 
is defined in the $\msbar$ scheme with five active (massless) flavors. 
The scaling functions $f_{ij}$ are proportional to the 
partonic cross sections and can be 
expanded in powers of $\alpha_s/\pi$.  Numerical results 
for the next-to-leading order (NLO) term
have been known for over two decades \cite{Nason:1987xz, Beenakker:1988bq,
  Beenakker:1990maa}, and more recently fully analytic results were obtained in
\cite{Czakon:2008ii}. Predictions based on NLO calculations typically exhibit scale
uncertainties larger than 10\%. In order to further reduce the theoretical uncertainties
to match the experimental precision, it is necessary to go beyond NLO. Consequently, the
calculation of the next-to-next-to-leading order (NNLO) corrections is an active area of
research. Many efforts have been made in the calculation of two-loop virtual corrections
\cite{Czakon:2007ej, Czakon:2007wk, Czakon:2008zk, Bonciani:2008az, Bonciani:2009nb,
  Bonciani:2010mn}, one-loop interference terms \cite{Korner:2008bn, Anastasiou:2008vd,
  Kniehl:2008fd}, and double real emission \cite{Dittmaier:2007wz, Dittmaier:2008yq,
  Bevilacqua:2010ve, Melnikov:2010iu}. Due to very recent progress in developing a new
subtraction scheme for double real emission in the presence of massive particles
\cite{Czakon:2010td, Czakon:2011ve, Anastasiou:2010pw, Abelof:2011jv, Bernreuther:2011jt},
a calculation of the NNLO corrections now seems feasible, and its completion
will be a major accomplishment.

An important way to improve on the fixed-order calculations is to supplement them with
threshold resummation \cite{Sterman:1986aj, Catani:1989ne}. In such an approach, which in
general works at the level of differential cross sections, one identifies a partonic
threshold parameter which vanishes in the limit where extra real radiation is soft, and
sums a certain tower of logarithmic corrections in this parameter to all orders in the
strong coupling. Such resummed formulas neglect power corrections which vanish when the
threshold parameter goes to zero, but these can be taken into account, up to a given order
in the strong coupling, by matching with the fixed-order results. In this way, one obtains
predictions which not only make full use of the fixed-order calculations, but also resum a
class of logarithmic corrections to all orders. In the limit where the higher-order
corrections are dominated by these logarithmic terms, such a resummation is clearly an
improvement. For this reason, many different implementations of threshold resummation have
been considered in the literature, where the current frontier is next-to-next-to-leading
logarithmic (NNLL) accuracy \cite{Moch:2008qy, Langenfeld:2009wd, Czakon:2009zw,
  Beneke:2009ye, Ahrens:2009uz, Ahrens:2010zv, Kidonakis:2010dk, Ahrens:2011mw}.

The purpose of this letter is to consolidate results for the total cross section based on
threshold resummation in effective field theory. These are obtained from two different
threshold limits for the differential cross section. The first, referred to as
pair invariant-mass (PIM) kinematics, uses the top-pair invariant-mass distribution as the
fundamental observable \cite{Ahrens:2009uz, Ahrens:2010zv}. The second, referred to as
single-particle inclusive (1PI) kinematics, works at the level of the transverse-momentum
or rapidity distributions of the top quark \cite{Ahrens:2011mw}. The total partonic cross
section is obtained by integrating the resummed distributions over the appropriate phase
space. In both kinematics, the predictions in the effective field-theory framework include 
resummation effects to NNLL order, and are matched with the fixed-order results at
NLO in order to achieve NLO+NNLL accuracy for the total cross section. Such formulas can
be evaluated numerically using specific values of the matching scales appearing in the
effective-theory calculations, or otherwise re-expanded in a fixed-order expansion, in the
form of approximate NNLO predictions. We have considered both scenarios in
\cite{Ahrens:2010zv, Ahrens:2011mw}, finding that when the results in PIM and 1PI kinematics 
are combined as in the present work, the numerical differences between the NLO+NNLL and
approximate NNLO results are rather small. In this letter we focus on the approximate NNLO
results for concreteness.

In what follows, we briefly review the formalism, and then
give our best numerical predictions for the total cross section in
the pole scheme for the top-quark mass, in the form of numerical fits as functions of
$m_t$.  We then explain how to convert the results to the $\msbar$ scheme and present
numerical results as a function of the $\msbar$ mass. The results in both schemes are 
calculated with a Fortran implementation of our approximate NNLO formulas, which we
include with the electronic submission of this letter.  Contrary to
\cite{Langenfeld:2009wd}, we do not find a strong improvement of the perturbative series
in the $\msbar$ scheme compared to the pole scheme. 
We suggest that the group of short-distance masses known as threshold masses
(see e.g. \cite{Hoang:2000yr}) may actually be the more appropriate choice,
and we explicitly consider the case of the mass defined in the 1S scheme 
\cite{Hoang:1999zc}. However, we believe that the poor large-order behavior of the 
perturbative series in the pole
scheme is unlikely to be of practical importance in the foreseeable future.

\section{Ingredients of the calculation}

Our calculation is based on \cite{Ahrens:2010zv, Ahrens:2011mw}, where threshold
resummation using soft-collinear effective theory (SCET) \cite{Bauer:2000yr, Bauer:2001yt,
  Beneke:2002ph} is applied to $t\bar{t}$ production at hadron colliders. The partonic
scattering process is
\begin{align}
  i(p_1) + j(p_2) \to t(p_3) + \bar{t}(p_4) + \hat{X}(k) \, ,
\end{align}
where $i$, $j$ indicate the incoming partons and $\hat{X}$ is a partonic final state. In
\cite{Ahrens:2010zv}, the invariant-mass distribution $d\sigma/dM$ of the $t\bar{t}$ pair was
considered, where $M^2 \equiv (p_3+p_4)^2$, and the threshold region is defined by
$z=M^2/\hat{s} \to 1$ with $\hat{s}=(p_1+p_2)^2$ (so-called PIM kinematics). In
\cite{Ahrens:2011mw}, on the other hand, the transverse-momentum and rapidity
distributions of the top-quark were considered. In this case,
the threshold region is defined by $s_4=(p_4+k)^2-m_t^2 \to 0$ (1PI kinematics).
In both the PIM and 1PI threshold limits, the energy of the extra emitted gluons vanishes,
and the partonic differential cross sections are dominated by singular distributions of
the form
\begin{align}
  \alpha_s^n \left[ \frac{\ln^m\xi}{\xi} \right]_+  ; \quad m=0,\ldots,2n-1 \, ,
\end{align}
where $\xi=(1-z)/\sqrt{z}$ for PIM and $\xi=s_4/(m_t\sqrt{m_t^2+s_4})$ for 1PI kinematics. 

In the threshold limit $\xi \to 0$, the partonic differential cross section can be
factorized into a product of matrix-valued hard and soft functions,
\begin{align}
  \label{eq:partfact}
  d\hat{\sigma}(\xi,\mu_f) \propto \mathrm{Tr} \left[ \bm{H}(\mu_f) \, \bm{S} \bigg(
      \frac{E_s(\xi)}{\mu_f}, \alpha_s(\mu_f) \bigg) \right] ,
\end{align}
where we have suppressed the dependence on the other kinematic variables, and $E_s(\xi)$
is the energy of the soft gluon radiation, which is given by $E_s=M\xi$ in the partonic
center-of-mass frame for PIM kinematics, and $E_s=m_t\,\xi$ in the $\bar{t}$ rest frame 
for 1PI kinematics. The
hard function $\bm{H}$ is related to virtual corrections and is thus evaluated at
$\xi=0$. The $\xi$ dependence of the cross section is encoded in the soft function
$\bm{S}$ via the ratio $E_s(\xi)/\mu_f$. One can therefore use the renormalization-group
(RG) equation of the soft function to resum the singular distributions in $\xi$ to
all orders in $\alpha_s$. A technical complication is that the RG equation of the soft
function is non-local. We solve this equation using the technique of Laplace transforms
\cite{Becher:2006nr}, and then carry out the inverse Laplace transform analytically to
obtain expressions in momentum space. The final result for the resummed cross section reads
\begin{align}
  d\hat{\sigma}(\xi,\mu_f) &\propto \exp \bigl[ 4a_{\gamma^\phi}(\mu_s,\mu_f) \bigr] \,
  \mathrm{Tr} \Bigl[ \bm{U}(\mu_h,\mu_s) \, \bm{H}(\mu_h) \, \bm{U}^\dagger(\mu_h,\mu_s)
  \, \tilde{\bm{s}}(\partial_\eta,\alpha_s(\mu_s)) \Bigr] \,
  \frac{e^{-2\gamma_E\eta}}{\Gamma(2\eta)} \, \xi^{-1+2\eta} \, ,
\end{align}
where $\tilde{\bm{s}}$ is the Laplace transform of the soft function, $\bm{U}$ is an
evolution matrix which resums large logarithms between the hard and soft scales $\mu_h$ and 
$\mu_s$, and $\eta=2a_\Gamma(\mu_s,\mu_f)$ and $a_{\gamma^\phi}(\mu_s,\mu_f)$ are related to 
certain integrals over anomalous-dimension functions. Explicit expressions for these objects 
in RG-improved perturbation theory can be
found in \cite{Ahrens:2010zv}. For NNLL accuracy, one needs the anomalous-dimension
matrices for the hard and soft functions to two-loop order (computed in
\cite{Ferroglia:2009ii}), as well as the hard and soft functions 
to one-loop order (computed in \cite{Ahrens:2010zv, Ahrens:2011mw}). One must also specify
a procedure for choosing the hard and soft scales. In the SCET approach, these are chosen
such that the contributions of the NLO matching functions to the hadronic cross section
are minimized; details can be found in \cite{Ahrens:2010zv, Ahrens:2011mw}.

The effective theory is a powerful tool for separating physics at the
formally very different scales $\mu_h$ and $\mu_s$ and for summing
logarithms of their ratio.  In practice, however, in areas of phase
space where the differential cross section is largest, the numerical
hierarchy between the two scales is only moderate.  Therefore, an
alternative to using the NLO+NNLL formulas directly is to re-expand
them in a fixed-order series in $\alpha_s(\mu_f)$, 
constructing what are referred to as approximate NNLO formulas. The
full NNLO correction to the cross section takes the form
\begin{align}
  d\hat{\sigma}^{(2)}(\xi,\mu_f) &\propto \alpha_s^4(\mu_f)
  \left( \sum \limits_{m = 0}^3 D_m \left[\frac{\ln^m\xi}{\xi} \right]_+ + C_0 \,
    \delta(\xi) + R(\xi) \right) . \label{eq:ds2}
\end{align}
The approximate NNLO formulas derived from the NLO+NNLL results contain an unambiguous
answer for the $D_m$ coefficients in the limit $\xi\to 0$. Only parts of the coefficient $C_0$ are
determined, and the regular piece $R$ is subleading in the threshold limit and not
determined. There are thus ambiguities in what to include in the $C_0$ term, and a freedom
to shuffle terms between the $D_i$ and $R$ away from the exact limit $\xi \to 0$. For our
method of dealing with these ambiguities, we refer the reader to \cite{Ahrens:2011mw},
which specifies all details relevant for our implementation of the approximate NNLO
formulas used in the numerical analysis.

Equations~(\ref{eq:ff}) and (\ref{eq:partfact})--(\ref{eq:ds2}) are written for the case where 
the factorization 
and renormalization scales are set equal, $\mu_f = \mu_r$. Later on we will consider the
case where the running coupling is instead evaluated at an arbitrary renormalization scale
$\mu_r$. To derive such expressions we use
\begin{align}
  \alpha_s(\mu_f) = \alpha_s(\mu_r) \left[ 1 + \frac{\alpha_s(\mu_r)}{\pi}
    \frac{23}{12}L_{rf} + \left(\frac{\alpha_s(\mu_r)}{\pi}\right)^2
    \left(\frac{529}{144}L_{rf}^2 + \frac{29}{12}L_{rf}\right) \right] ,
\end{align}
with $L_{rf} = \ln(\mu_r^2/\mu_f^2)$, and re-expand the formulas to NNLO in powers of
$\alpha_s(\mu_r)$.

Although PIM and 1PI kinematics are used to describe different differential distributions,
one can always integrate over the distributions to obtain the total hadronic cross section
or the partonic scaling functions $f_{ij}$ using the two approaches. Since the two
kinematics both account for soft gluon emission, they give rise to the same results for
$f_{ij}$ in the limit $\hat{s} \to 4m_t^2$. When $\hat{s}$ deviates from $4m_t^2$, the
difference between the two kinematics is formally subleading in $\xi$. However, the
numerical differences may be visible or even significant if $\hat{s}$ is much larger than
$4m_t^2$. In \cite{Ahrens:2011mw}, we have shown that the effective-theory predictions for
the total cross section from the two kinematics actually agree quite well, as long as the
exact dependence on the energy of soft gluon radiation is kept in the factorization
formula (\ref{eq:partfact}).\footnote{Previous calculations expanded $E_s(\xi)$ in the
  limit $z\to 1$ for PIM kinematics and $s_4\to 0$ for 1PI kinematics, in which case the
  differences between the two types of kinematics become much larger.} 
The differences between
these two kinematics can be regarded as another source of theoretical uncertainty, namely
that due to power corrections to the soft limit, and used along with scale uncertainties
to estimate the total uncertainty associated with the calculation of the total cross section
at NLO+NNLL or approximate NNLO. In the following sections we define our procedure
for combining the two types of uncertainties and give our final results for the total
cross section as a function of the top-quark mass in the pole, $\msbar$, and 1S schemes.

\section{Cross sections in the pole scheme}

We now specify our method for estimating the theoretical uncertainties from scale variations, PDF variations and variations of $\alpha_s$ in the calculation of the top-pair production cross section, using either PIM or 1PI kinematics. We then define a procedure for combining the results obtained in these two schemes. By default, the pole mass $m_t$ is used in the calculation of the partonic cross sections. We comment later on alternative schemes for defining the top-quark mass.

To estimate the uncertainties associated with scale variations,  we view the cross 
section as a function of the renormalization and factorization scales, which by 
default are chosen as $\mu_f = \mu_r = m_t$.  We then consider two methods
of scale variations: correlated variations with $\mu_f = \mu_r$ 
varied up and down by
a factor of two from the default value, and independent variations of $\mu_f$ and $\mu_r$
by factors of two, with the uncertainties added in quadrature. We use as our final
answer the larger uncertainty from these two methods. 

To combine the results from PIM and 1PI kinematics, we first compute the cross sections
and scale uncertainties in the PIM and 1PI schemes separately, and obtain six quantities
$\sigma_{\text{PIM}}$, $\Delta\sigma_{\text{PIM}}^+$, $\Delta\sigma_{\text{PIM}}^-$,
$\sigma_{\text{1PI}}$, $\Delta\sigma_{\text{1PI}}^+$, $\Delta\sigma_{\text{1PI}}^-$. The
central value and perturbative uncertainties for the combined results are then
determined by
\begin{align}
  \sigma &= \frac12 \left( \sigma_{\text{PIM}} + \sigma_{\text{1PI}} \right)  ,
  \nonumber\\
  \Delta\sigma^+ &= {\rm max} \left( \sigma_{\text{PIM}}+\Delta\sigma_{\text{PIM}}^+,
    \sigma_{\text{1PI}}+\Delta\sigma_{\text{1PI}}^+ \right) - \sigma \, , 
  \\
  \Delta\sigma^- &= {\rm min} \left( \sigma_{\text{PIM}}+\Delta\sigma_{\text{PIM}}^-,
    \sigma_{\text{1PI}}+\Delta\sigma_{\text{1PI}}^- \right) - \sigma \, . \nonumber
\end{align}
In this way, the central value is the average of the two, and the perturbative
uncertainties reflect both the variation of the scales and the difference between the
two types of kinematics. The PDF uncertainties are estimated as usual by evaluating the
average of the 1PI and PIM results using the PDF error sets at a particular confidence level.

We quote in Table~\ref{tab:best} the approximate NNLO predictions obtained with the above
procedure at $m_t=173.1$~GeV, using the 90\% CL sets of the MSTW2008 NNLO PDFs
\cite{Martin:2009iq}. To investigate the convergence of the perturbative series, we also
list the LO and NLO results, obtained using MSTW2008 LO and NLO PDFs,
respectively.  In the 
pole scheme, the scale
uncertainties are generally determined by the correlated scale variations with $\mu_r=\mu_f$.
The one exception is the upper error at the Tevatron, which is instead determined by the
independent variations of $\mu_r$ and $\mu_f$ added in quadrature. Even though the
perturbative uncertainty in the approximate NNLO result includes both scale variations and
an estimate of power corrections to the soft limit through the difference of 1PI and PIM
kinematics, it is still reduced compared to that in the NLO calculation, which by
definition is due only to scale variations.  
We note that the central value and uncertainties of the approximate NNLO results 
are well contained within the uncertainty  range predicted by the NLO results, 
so that the perturbative series to this order is well behaved in the pole scheme.
The NNLO results are also within the uncertainties of the LO calculation, 
although the NLO results are slightly higher than the LO ones 
in the case of the LHC.   

\begin{table}[h]
  \centering
  \begin{tabular}{|r|c|c|c|c|c|c|c|cc|c|}
    \hline
    &  \multicolumn{2}{c|}{Tevatron} & \multicolumn{2}{c|}{LHC7} & \multicolumn{2}{c|}{LHC14}
    \\ \cline{2-7}
    & MSTW & CTEQ & MSTW & CTEQ &  MSTW & CTEQ 
    \\ \hline
    LO &
    $6.66^{+2.95+(0.34)}_{-1.87-(0.27)}$ & $5.45^{+2.16+0.33(0.29)}_{-1.42-0.27(0.24)}$ &
    $122^{+49+(6)}_{-32-(7)}$ & $100^{+35+9(7)}_{-24-8(7)}$ &
    $681^{+228+(26)}_{-159-(34)}$ & $552^{+157+25(18)}_{-115-25(19)}$
    \\ \hline
    NLO &
    $6.72^{+0.41+0.47(0.37)}_{-0.76-0.45(0.24)}$ & $6.77^{+0.40+0.50(0.43)}_{-0.74-0.40(0.34)}$& 
    $159^{+20+14(8)}_{-21-13(9)}$ & $148^{+18+13(11)}_{-19-12(10)}$ &
    $889^{+107+66(31)}_{-106-58(32)}$ & $829^{+97+41(27)}_{-96-40(28)}$
    \\ \hline
    NNLO approx. &
    $6.63^{+0.07+0.63(0.33)}_{-0.41-0.48(0.25)}$ & $6.91^{+0.09+0.53(0.46)}_{-0.44-0.43(0.36)}$ &
    $155^{+8+14(8)}_{-9-14(9)}$ & $153^{+8+13(11)}_{-8-12(10)}$ &
    $855^{+52+60(30)}_{-38-59(31)}$ & $842^{+51+40(26)}_{-37-40(28)}$
    \\ \hline
  \end{tabular}
  \caption{Total cross sections in pb for $m_t=173.1$~GeV with MSTW2008 and CTEQ6.6 PDFs. The first
    error results from the perturbative uncertainty from both scale variations and the
    difference between PIM and 1PI kinematics, the second one accounts for the combined PDFs+$\alpha_s$ 
    uncertainty. The numbers in parenthesis show the PDF uncertainty only.}
  \label{tab:best}
\end{table}

For comparison, we also include the results using CTEQ6.6 PDFs \cite{Nadolsky:2008zw} in
Table~\ref{tab:best}. Since the CTEQ PDFs are based on a NLO fit, the same
set is used at LO, NLO and approximate NNLO. The statements based on
the analysis with MSTW PDFs above, including those 
concerning the moderate size of the NNLO corrections, are also true for the 
analysis with CTEQ PDFs. In this case, however, the LO results at the LHC are 
significantly lower than the NLO and NNLO results.  To a certain extent, this 
shows the potential benefit of switching PDFs as appropriate to the order 
of perturbation theory. On the other hand, LO calculations are usually considered
unreliable, so the more important observation for the perturbative convergence
is the modest size of the NNLO correction.   

The perturbative uncertainties in the approximate NNLO predictions are about the same size
at both the Tevatron and the LHC. An additional source of uncertainty is related to the
experimental value of $\alpha_s(M_Z)$ (where $M_Z$ denotes the $Z$-boson mass), which is
an input parameter for the running of the strong coupling constant. We estimate this
uncertainty in combination with the PDF one by employing the method proposed in
\cite{Martin:2009bu, Lai:2010nw}. Table~\ref{tab:best} shows that the uncertainty on
$\alpha_s(M_Z)$ adds an error of $\pm(3\mbox{\,--\,}4)\%$ to the pair-production cross
section when the calculation is carried out with MSTW2008 PDFs. The error is somewhat
smaller, $\pm(1\mbox{\,--\,}2)\%$, when CTEQ6.6 PDFs are used. The reason is that CTEQ6.6
assigns a 90\% CL error of $\pm 0.002$ to $\alpha_s(M_Z)$, while for MSTW2008 it is $\pm
0.003$. One can conclude that the $\alpha_s(M_Z)$ induced uncertainty is of the same order
of magnitude as the perturbative and PDF uncertainties, and should not be neglected.

For an extraction of the top-quark mass through a comparison with the experimental cross
section, we also provide our results as a function of $m_t$. We parametrize the mass
dependence of the approximate NNLO cross section using the simple polynomial fit
\begin{align}
  \label{eq:fit}
  \sigma(m_t) = c_0 + c_1 x + c_2 x^2 + c_3 x^3 + c_4 x^4 \, ,
\end{align}
where $x=m_t/\text{GeV}-173$, and $c_i$ are fit coefficients which depend on the collider
and the PDF set. The results for the fit coefficients including upper and lower errors due
to perturbative uncertainties are shown in Table~\ref{tab:fit}, again using MSTW2008 NNLO
PDFs. A {\tt Mathematica} implementation of the fit coefficients can be found with the
electronic version of this letter, where the combined PDF and $\alpha_s$ uncertainties as 
well as the fit coefficients using CTEQ6.6 PDFs are also included. These fits reproduce the
approximate NNLO calculations to 1 permille or better in the range $m_t \in
[150,180]$~GeV. For simplicity the uncertainties on fit coefficients are not displayed
in Table~\ref{tab:fit}. When these uncertainties are measured in percent of
the central value of the cross section, they appear to be roughly independent of $m_t$ in
the range $m_t \in [150,180]$~GeV, differing by no more than a percent from those at
$m_t=173.1$~GeV shown in Table~\ref{tab:best}.

\begin{table}[h]
  \begin{tabular}{|c|c|c|c|c|c|c|}
    \hline
    & & $c_0$~[pb] & $c_1$~[pb] & $c_2$~[pb] & $c_3$~[pb] & $c_4$~[pb]
    \\ \hline
    Tevatron & $\sigma$ &
    $6.64792 \times 10^{0}$ & $-2.07262 \times 10^{-1}$ & $3.61739 \times 10^{-3}$ &
    $-4.30451 \times 10^{-5}$ & $8.94347 \times 10^{-7}$
    \\
     & $\sigma+\Delta\sigma^+$ &
    $6.72257 \times 10^{0}$ & $-2.09199 \times 10^{-1}$ & $3.62959 \times 10^{-3}$ &
    $-5.00960 \times 10^{-5}$ & $6.99427 \times 10^{-7}$
    \\
     & $\sigma+\Delta\sigma^-$ &
    $6.23323 \times 10^{0}$ & $-1.94555 \times 10^{-1}$ & $3.40149 \times 10^{-3}$ &
    $-4.03465 \times 10^{-5}$ & $8.17661 \times 10^{-7}$
    \\ \hline
    LHC7 & $\sigma$ &
    $1.55546 \times 10^{2}$ & $-4.66554 \times 10^{0}$ & $8.07632 \times 10^{-2}$ &
    $-9.93138 \times 10^{-4}$ & $1.75303 \times 10^{-5}$
    \\
     & $\sigma+\Delta\sigma^+$ &
    $1.63325 \times 10^{2}$ & $-4.92409 \times 10^{0}$ & $8.50408 \times 10^{-2}$ &
    $-1.13686 \times 10^{-3}$ & $1.66808 \times 10^{-5}$
    \\
     & $\sigma+\Delta\sigma^-$ &
    $1.46870 \times 10^{2}$ & $-4.40296 \times 10^{0}$ & $7.51759 \times 10^{-2}$ &
    $-1.03349 \times 10^{-3}$ & $1.48145 \times 10^{-5}$
    \\ \hline
    LHC14 & $\sigma$ &
    $8.57671 \times 10^{2}$ & $-2.29929 \times 10^{1}$ & $3.65310 \times 10^{-1}$ &
    $-4.02700 \times 10^{-3}$ & $7.41522 \times 10^{-5}$
    \\
     & $\sigma+\Delta\sigma^+$ &
    $9.08856 \times 10^{2}$ & $-2.44668 \times 10^{1}$ & $3.89564 \times 10^{-1}$ &
    $-4.48258 \times 10^{-3}$ & $8.14287 \times 10^{-5}$
    \\
     & $\sigma+\Delta\sigma^-$ &
    $8.19550 \times 10^{2}$ & $-2.20396 \times 10^{1}$ & $3.54379 \times 10^{-1}$ &
    $-3.81968 \times 10^{-3}$ & $6.25216 \times 10^{-5}$
    \\ \hline
  \end{tabular}
  \caption{\label{tab:fit} Fit coefficients in (\ref{eq:fit}) for the total 
    cross sections with perturbative uncertainties at approximate NNLO, using  
    MSTW2008 NNLO PDFs.}
\end{table}

\section{Cross sections in the $\msbar$ and 1S schemes}

It is well-known that the pole mass of a quark cannot be defined unambiguously in QCD due to
confinement;  the perturbatively defined pole mass is sensitive to long-distance
physics and suffers from renormalon ambiguities of order $\Lambda_{\text{QCD}}$
\cite{Bigi:1994em, Beneke:1994sw}. In perturbative calculations, the renormalon 
ambiguity is associated with large higher-order corrections to the pole mass, 
and thus to any observable calculated in this scheme. 
Therefore, it is worth investigating short-distance mass definitions which are free 
from these shortcomings. In this section, we analyze the cross section as a function 
of the running top-quark mass defined in the $\msbar$ scheme, and of the threshold
top-quark mass defined in the 1S scheme \cite{Hoang:1999zc}.  

It is possible to calculate the cross section using the $\msbar$ mass
from the beginning, by performing mass renormalization in that scheme.
However, since we already have the cross section in the pole scheme,
it is simpler to convert from one scheme to another using
the perturbative relation between the pole mass
and $\msbar$ mass.  This relation is 
currently known to three-loop order \cite{Chetyrkin:1999ys}.  To 
perform the conversion to the $\msbar$ scheme, we take that result for 
QCD with five active flavors and write it in the form
\begin{align}
  \label{eq:conversion}
  m_t = \mbar(\mubar) \left[ 1 + \frac{\alpha_s(\mu_r)}{\pi} \, d^{(1)} +
    \frac{\alpha_s^2(\mu_r)}{\pi^2} \, d^{(2)} + \mathcal{O}(\alpha_s^3) \right] ,
\end{align}
where
\begin{align}
  d^{(1)} = \frac{4}{3} + L_m \, , \qquad
  d^{(2)} = 8.23656 + \frac{379}{72} L_m + \frac{37}{24} L_m^2 + \frac{23}{12} \, d^{(1)}
  L_r \, ,
\end{align}
with $L_m=\ln(\mubar^2/\mbar^2(\mubar))$ and $L_r=\ln(\mu_r^2/\mubar^2)$. We then decompose 
the NNLO cross section in the pole scheme as
\begin{align}
  \sigma_{\text{NNLO}}(m_t) = \left[ \frac{\alpha_s(\mu_r)}{\pi} \right]^2
   \sigma^{(0)}(m_t,\mu_r) +
  \left[ \frac{\alpha_s(\mu_r)}{\pi} \right]^3 \sigma^{(1)}(m_t,\mu_r) +
  \left[ \frac{\alpha_s(\mu_r)}{\pi} \right]^4 \sigma^{(2)}(m_t,\mu_r) \, ,
\end{align}
eliminate $m_t$ through the relation (\ref{eq:conversion}), and re-expand the result in powers of
$\alpha_s(\mu_r)$. The resulting cross section in the $\msbar$ scheme can be written as
\begin{align}
  \bar{\sigma}_{\text{NNLO}}(\mbar) 
  = \left[ \frac{\alpha_s(\mu_r)}{\pi} \right]^2
  \bar{\sigma}^{(0)}(\mbar(\mubar),\mubar,\mu_r) 
  + \left[ \frac{\alpha_s(\mu_r)}{\pi} \right]^3
  \bar{\sigma}^{(1)}(\mbar(\mubar),\mubar,\mu_r) 
  + \left[ \frac{\alpha_s(\mu_r)}{\pi} \right]^4
  \bar{\sigma}^{(2)}(\mbar(\mubar),\mubar,\mu_r) \, ,
\end{align}
where
\begin{align}
\label{eq:msbarconv}
  \bar{\sigma}^{(0)}(\mbar(\mubar),\mubar,\mu_r) &= \sigma^{(0)}(\mbar(\mubar),\mu_r) \, ,
  \nonumber
  \\
  \bar{\sigma}^{(1)}(\mbar(\mubar),\mubar,\mu_r) &= \sigma^{(1)}(\mbar(\mubar),\mu_r) +
  \mbar(\mubar) \, d^{(1)} \left[ \frac{d\sigma^{(0)}(m_t,\mu_r)}{dm_t}
  \right]_{m_t=\mbar(\mubar)} ,
  \\
  \bar{\sigma}^{(2)}(\mbar(\mubar),\mubar,\mu_r) &= \sigma^{(2)}(\mbar(\mubar),\mu_r)
  \nonumber
  \\
  &\quad + \mbar(\mubar) \left[ d^{(1)} \frac{d\sigma^{(1)}(m_t,\mu_r)}{dm_t} + d^{(2)}
    \frac{d\sigma^{(0)}(m_t,\mu_r)}{dm_t} + \frac{\left(d^{(1)}\right)^2\mbar(\mubar)}{2}
    \frac{d^2\sigma^{(0)}(m_t,\mu_r)}{dm_t^2} \right]_{m_t=\mbar(\mubar)} . \nonumber
\end{align}
The derivatives can be taken either at the level of the hadronic cross section, using fits
such as the one in (\ref{eq:fit}), or at the level of the differential cross section before 
carrying out the phase-space integrations. We have checked our calculations by verifying the
agreement between the two methods.  We note that our method of converting results from 
the pole scheme to the $\msbar$ scheme is similar to that used in 
\cite{Langenfeld:2009wd, Aliev:2010zk}.  Indeed, our approximate NNLO results in the $\msbar$
scheme for the choice $\mubar = \mbar$ agree with those in the HATHOR program
\cite{Aliev:2010zk}, apart from 
the piece related to the NNLO correction $\sigma^{(2)}$, which is of course different since
we are not working in the $\hat{s} \to 4m_t^2$ limit of the partonic cross section.

Our procedure for combining the results from 1PI and PIM kinematics in
the $\msbar$ scheme is analagous to that for the pole scheme described
above. In the present case, we use by default $\mu_f = \mu_r = \mbar(\mbar)$.
We must also specify the scale in the running top-quark mass, for which we use 
$\mubar=\mbar$.\footnote{Variations of $\mubar$ around values
close to $\mbar$, which would correspond to sampling over different mass definitions, 
could potentially be used as an additional
means of estimating systematic uncertainties. However, a numerical
analysis shows that our approximate NNLO results are very
stable for variations of $\bar\mu$ around the default value.}   
We provide results  for the cross sections as a function of
$\mbar(\mbar)$ using the fit
\begin{equation}
  \label{eq:msbarfit}
  \sigma(\mbar) = \bar{c}_0 + \bar{c}_1 \bar{x} + \bar{c}_2 \bar{x}^2 
  + \bar{c}_3 \bar{x}^3 + \bar{c}_4 \bar{x}^4 \, ,
\end{equation}
where $\bar{x}=\mbar/\text{GeV}-164$. The fit coefficients for the different colliders
using the MSTW2008 NNLO PDFs can be found in Table~\ref{tab:msbarfit}; those including
combined PDF and $\alpha_s$ uncertainties also with CTEQ6.6 PDFs are included in the 
{\tt Mathematica} notebook mentioned above.

\begin{table}[h]
  \begin{tabular}{|c|c|c|c|c|c|c|}
    \hline
    & & $\bar c_0$~[pb] & $\bar c_1$~[pb] & $\bar c_2$~[pb] & $\bar c_3$~[pb] & $\bar c_4$~[pb]
    \\ \hline
    Tevatron & $\sigma$ &
    $6.66715 \times 10^{0}$ & $-2.17800 \times 10^{-1}$ & $3.95994 \times 10^{-3}$ &
    $-5.14404 \times 10^{-5}$ & $1.09983 \times 10^{-6}$
    \\
     & $\sigma+\Delta\sigma^+$ &
    $6.77748 \times 10^{0}$ & $-2.21151 \times 10^{-1}$ & $4.04717 \times 10^{-3}$ &
    $-5.09432 \times 10^{-5}$ & $1.11678 \times 10^{-6}$
    \\
     & $\sigma+\Delta\sigma^-$ &
    $6.26205 \times 10^{0}$ & $-2.05259 \times 10^{-1}$ & $3.81108 \times 10^{-3}$ &
    $-3.88796 \times 10^{-5}$ & $1.28549 \times 10^{-6}$
    \\ \hline
    LHC7 & $\sigma$ &
    $1.57441 \times 10^{2}$ & $-4.94191 \times 10^{0}$ & $9.00990 \times 10^{-2}$ &
    $-9.38583 \times 10^{-4}$ & $2.97762 \times 10^{-5}$
    \\
     & $\sigma+\Delta\sigma^+$ &
    $1.66413 \times 10^{2}$ & $-5.20036 \times 10^{0}$ & $9.48216 \times 10^{-2}$ &
    $-1.04597 \times 10^{-3}$ & $2.81345 \times 10^{-5}$
    \\
     & $\sigma+\Delta\sigma^-$ &
    $1.48389 \times 10^{2}$ & $-4.62721 \times 10^{0}$ & $8.26628 \times 10^{-2}$ &
    $-1.06324 \times 10^{-3}$ & $2.22025 \times 10^{-5}$
    \\ \hline
    LHC14 & $\sigma$ &
    $8.64542 \times 10^{2}$ & $-2.42364 \times 10^{1}$ & $3.98093 \times 10^{-1}$ &
    $-4.89960 \times 10^{-3}$ & $8.44709 \times 10^{-5}$
    \\
     & $\sigma+\Delta\sigma^+$ &
    $9.20794 \times 10^{2}$ & $-2.59880 \times 10^{1}$ & $4.34218 \times 10^{-1}$ &
    $-5.00638 \times 10^{-3}$ & $1.02994 \times 10^{-4}$
    \\
     & $\sigma+\Delta\sigma^-$ &
    $8.03504 \times 10^{2}$ & $-2.21450 \times 10^{1}$ & $3.60816 \times 10^{-1}$ &
    $-4.19956 \times 10^{-3}$ & $7.84995 \times 10^{-5}$
    \\ \hline
  \end{tabular}
  \caption{\label{tab:msbarfit} Fit coefficients (\ref{eq:msbarfit}) 
    for the cross section with perturbative uncertainties at approximate NNLO 
    in the $\msbar$ scheme, using MSTW2008 NNLO PDFs.}
\end{table}

The results for $\mbar(\mbar) = 164.1$~GeV, which corresponds to $m_t = 173.1$~GeV when
using the two-loop conversion between the pole and $\msbar$ masses, are shown in
Table~\ref{tab:msbarbest} for MSTW2008 and CTEQ6.6 PDFs. As in the pole scheme, we switch
the order of the MSTW PDFs according to the order of perturbation theory at which we are
working, while the CTEQ PDFs are the same in both cases. In the $\msbar$ scheme, the
uncertainties from scale variations are dominated by the scheme where $\mu_f$ and $\mu_r$
are varied independently, rather than the scheme with
correlated $\mu_r = \mu_f$ variations, as was the case in the pole scheme.

\begin{table}[h]
  \centering
  \begin{tabular}{|r|c|c|c|c|c|c|c|cc|c|}
    \hline
    &  \multicolumn{2}{c|}{Tevatron} & \multicolumn{2}{c|}{LHC7} & \multicolumn{2}{c|}{LHC14}
    \\ \cline{2-7}
    & MSTW & CTEQ & MSTW & CTEQ &  MSTW & CTEQ 
    \\ \hline
    LO &
    $8.82^{+3.91+(0.44)}_{-2.48-(0.35)}$ & $7.24^{+2.86+0.46(0.40)}_{-1.89-0.38(0.32)}$ &
    $160^{+64+(8)}_{-42-(9)}$ & $131^{+45+11(9)}_{-31-10(8)}$ &
    $875^{+291+(32)}_{-204-(43)}$ & $705^{+199+30(21)}_{-145-30(23)}$
    \\ \hline
    NLO &
    $7.33^{+0.11+0.50(0.40)}_{-0.49-0.47(0.25)}$ & $7.39^{+0.10+0.57(0.50)}_{-0.48-0.45(0.39)}$ &
    $179^{+11+15(10)}_{-19-14(10)}$ & $167^{+10+15(12)}_{-17-13(11)}$ &
    $991^{+79+71(35)}_{-96-62(36)}$ & $925^{+71+44(29)}_{-87-43(31)}$
    \\ \hline
    NNLO approx. &
    $6.64^{+0.11+0.58(0.33)}_{-0.40-0.43(0.23)}$ & $6.92^{+0.12+0.52(0.46)}_{-0.43-0.42(0.37)}$ &
    $157^{+9+13(8)}_{-9-13(9)}$ & $154^{+9+13(11)}_{-9-12(10)}$  &
    $862^{+56+54(30)}_{-61-53(32)}$ & $848^{+56+37(26)}_{-61-38(28)}$
    \\ 
    \hline
  \end{tabular}
  \caption{Total cross sections in pb in the $\msbar$ scheme, for $\mbar(\mbar) =164.1$~GeV. The first
    error results from the perturbative uncertainty from both scale variations and the
    difference between PIM and 1PI kinematics, the second one accounts for the combined PDFs+$\alpha_s$ 
    uncertainty. The numbers in parenthesis show the PDF uncertainty only.}
  \label{tab:msbarbest}
\end{table}

We observe that the results obtained from the approximate NNLO formulas are 
quite close to those in the pole scheme shown in Table~\ref{tab:best}, both in 
the central values and in the errors. Given this good agreement, which is 
roughly independent of the exact value of the top-quark mass as shown by
the fits, it makes little practical difference
whether one extracts the pole mass using the approximate NNLO results,
and then determines the $\msbar$ mass using the perturbative conversion
(\ref{eq:conversion}), or whether one determines the $\msbar$ mass directly, 
using the experimental results along with the fits at approximate NNLO.
This statement would not be true at very high orders in perturbation theory, since
the renormalon ambiguity inherent to the pole mass would lead to large corrections
not present in a short-distance scheme such as the $\msbar$ scheme. But
given the present accuracy of perturbative calculations and experimental
measurements, this does not yet appear to be an issue.  

It is of course still interesting to study whether even at low orders the perturbative
expansion is better behaved in the $\msbar$ scheme than in the pole scheme.
We observe that the perturbative uncertainties at NLO are generally 
smaller in the $\msbar$ scheme than in the pole scheme, and that the central 
values are relatively higher compared to the approximate NNLO calculation.  
For this reason, the overlap between the NLO and approximate NNLO results 
is actually better in the pole scheme than in the $\msbar$ scheme.\footnote{The 
overlap between LO and NLO is worse at the Tevatron and improved at the LHC compared
to the pole scheme, but as mentioned earlier we consider the more important
issue the overlap between the NLO and approximate NNLO results.}
These results differ from those obtained in the $\hat{s}\to 4m_t^2$ limit, 
where the approximated NNLO corrections and the perturbative
uncertainties at that order are significantly smaller in the $\msbar$ scheme 
than in the pole scheme \cite{Langenfeld:2009wd}. 

To elaborate further on these results, we note that the re-organization
of the perturbative expansion in the $\msbar$ scheme compared to the 
pole scheme is accomplished by the terms in square brackets in 
(\ref{eq:msbarconv}). 
To understand whether these terms are expected to cancel againt unphysically
large corrections in the pole scheme, we note that the 
main source of mass dependence in the 
Born level cross section is due to phase-space factors: 
the lower limit of integration 
in (\ref{eq:ff}), and an overall factor of $\sqrt{1-4m_t^2/\hat{s}}$ in the partonic 
cross section related to two-body phase space and multiplying the Born-level 
matrix element. The derivatives contained in the terms in square
brackets are mainly sensitive to those sources of $m_t$ dependence.   However, the phase
space of the pair production is more indicative of the pole
mass than of an $\msbar$ mass. Indeed, we are calculating the 
cross section for on-shell quarks according to the narrow width
approximation.   If the cross section is instead calculated in the $\msbar$ scheme, 
the terms in the square brackets of the NLO and NNLO pieces of (\ref{eq:msbarconv})
give sizeable negative corrections, which are accounted in the pole scheme by using
a numerically higher value of the mass in the LO and NLO cross sections. 
Since the most appropriate mass scheme for 
a given process is the one where the higher-order corrections are expected to be smallest
on physical grounds, it does not seem to us that the $\msbar$ scheme is the optimal
choice for this case.  

As an alternative to the $\msbar$ mass, we consider the group of short-distance masses
known as threshold masses \cite{Hoang:2000yr}. At lower orders in perturbation theory,
these are closer numerically to the pole mass, but they do not suffer from renormalon
ambiguities at higher orders. The cross section in these schemes can be easily calculated
from the pole-scheme results, using an analogous procedure to the $\msbar$ scheme
calculation. It is evident that at approximate NNLO the numerical difference between these
results and the $\msbar$ and pole-scheme results will be quite small once the numerical value 
of the mass is adjusted appropriately, but we nonetheless illustrate this with a specific
example. In particular, we consider the cross section as a function of the 1S mass
introduced in \cite{Hoang:1999zc}. The 1S mass is defined through the perturbative
contribution to the mass of a hypothetical $n=1$, $^3S_1$ toponium bound state. To perform
the conversion to this scheme, we write its relation with the pole mass in the form
\cite{Hoang:1999zc}
\begin{align}
  \label{eq:oneS}
  m_t = m_t^{\text{1S}} \left\{ 1 + \frac{\alpha_s(\mu_r)}{\pi} \, \frac{2}{9}
    \pi\alpha_s(\mu_r) + \left( \frac{\alpha_s(\mu_r)}{\pi} \right)^2 \left[ \frac{2}{9}
      \pi\alpha_s(\mu_r) \left( \frac{23}{3}
        \ln\frac{3\mu_r}{4\alpha_s(\mu_r)m_t^{\text{1S}}} + \frac{181}{18} + \frac{2}{9}
        \pi\alpha_s(\mu_r) \right) \right]+ \mathcal{O}\left(\frac{\alpha_s^3}{\pi^3}\right) \right\} ,
\end{align}
and follow the same procedure as for the $\msbar$ scheme calculation with the appropriate
replacements, cf.~(\ref{eq:conversion}). Note that in the above relation $\pi\alpha_s$ is counted as
  $\mathcal{O}(1)$ and is not expanded. The results are listed in Table~\ref{tab:1sbest}
for the value $m_t^{\rm 1S}=172.3$~GeV, which corresponds to a pole mass of
$m_t=173.1$~GeV using the two-loop conversion above. The approximate NNLO results in this
scheme are very similar to those in the pole and $\msbar$ schemes, but the moderate size
of the NNLO correction is more indicative of the pole scheme than of the $\msbar$ scheme.
This leads us to conclude once again that although at yet higher orders in perturbation
theory the pole mass would be disfavored, at approximate NNLO accuracy this is not yet a
problem.

\begin{table}[h]
  \centering
  \begin{tabular}{|r|c|c|c|c|c|c|c|cc|c|}
    \hline
    &  \multicolumn{2}{c|}{Tevatron} & \multicolumn{2}{c|}{LHC7} & \multicolumn{2}{c|}{LHC14}
    \\ \cline{2-7}
    & MSTW & CTEQ & MSTW & CTEQ &  MSTW & CTEQ 
    \\ \hline
    LO &
    $6.83^{+3.02+(0.35)}_{-1.92-(0.28)}$ & $5.59^{+2.21+0.34(0.30)}_{-1.46-0.28(0.24)}$ &
    $124^{+50+(6)}_{-33-(7)}$ & $103^{+35+9(7)}_{-24-8(7)}$ &
    $696^{+223+(26)}_{-163-(34)}$ & $564^{+160+25(18)}_{-117-25(19)}$
    \\ \hline
    NLO &
    $6.82^{+0.39+0.48(0.38)}_{-0.75-0.46(0.24)}$ & $6.87^{+0.38+0.51(0.44)}_{-0.73-0.41(0.34)}$ &
    $162^{+19+14(9)}_{-21-13(9)}$ & $150^{+17+14(11)}_{-19-12(10)}$ &
    $902^{+106+66(32)}_{-106-59(33)}$ & $841^{+96+41(27)}_{-96-40(29)}$
    \\ \hline
    NNLO approx. &
    $6.65^{+0.06+0.63(0.32)}_{-0.38-0.47(0.24)}$ & $6.93^{+0.08+0.54(0.47)}_{-0.40-0.42(0.36)}$ &
    $156^{+7+14(9)}_{-8-14(8)}$ & $154^{+7+13(11)}_{-8-12(10)}$  &
    $859^{+47+59(30)}_{-35-58(32)}$ & $846^{+46+39(25)}_{-35-40(29)}$
    \\ 
    \hline
  \end{tabular}
  \caption{Total cross sections in pb in the 1S scheme, for $m_t^{\rm 1S}=172.3$~GeV. The first
    error results from the perturbative uncertainty from both scale variations and the
    difference between PIM and 1PI kinematics, the second one accounts for the combined PDFs+$\alpha_s$ 
    uncertainty. The numbers in parenthesis show the PDF uncertainty only.}
  \label{tab:1sbest}
\end{table}

\section{Conclusions}

We have presented predictions for the total inclusive  cross section for top-quark 
pairs at hadron colliders at approximate NNLO in QCD. Our calculations are based 
on soft gluon resummation to NNLL order in PIM and 1PI kinematics, carried out within 
the context of effective field theory. They represent the state-of-the-art, combining 
all knowledge presently available about higher-order QCD corrections to the production 
cross section. The perturbative uncertainties associated
with our results are estimated in two ways: through the standard method of variations 
of factorization and renormalization scales, and also through the difference between the 
two types of kinematics.  The latter gives a means of estimating the size of 
perturbative power corrections to the soft limits in which the approximate NNLO
formulas are derived.  The results presented here consolidate those previously presented 
in \cite{Ahrens:2010zv, Ahrens:2011mw}.  We have also provided a computer program
which calculates the total cross section within our approach.  

The total production cross section can be used along with experimental measurements
to extract the top-quark mass.  An advantage of such extractions is that the theory
calculations are carried out in a well-defined renormalization scheme for $m_t$.  
For very precise extractions of the top-quark mass the pole mass is disfavored, because 
it is only defined up to a renormalon ambiguity of order $\Lambda_{\rm QCD}$. In practice, 
however, we have not observed a poor 
convergence of the perturbative series up to NNLO in the pole scheme compared to the 
$\msbar$ scheme, and pointed out that the group of short-distance masses known as threshold
masses may be equally appropriate.  We have provided numerical fits of our results  
as a function of the mass in both the pole and $\msbar$ schemes,  including perturbative
and PDF uncertainties, in addition to those from the strong coupling constant, which 
are non-negligible at this level of accuracy.  

The results presented in this letter can be used directly by the experimental
collaborations at the Tevatron and LHC in top-quark mass measurements from the corresponding 
production cross sections.  

\vspace{2mm}
{\em Acknowledgements:\/} 
This research was supported in part by the State of Rhineland-Palatinate via the Research 
Center {\em Elementary Forces and Mathematical Foundations}, by the German Federal Ministry 
for Education and Research under grant 05H09UME, by the German Research Foundation under 
grant NE398/3-1, 577401, by the European Commission through the {\em LHCPhenoNet\/} Initial 
Training Network PITN-GA-2010-264564, and by the Schweizer Nationalfonds under grant 
200020-124773.


\begin{thebibliography}{99}

\bibitem{Flacher:2008zq}
  H.~Fl\"acher, M.~Goebel, J.~Haller, A.~H\"ocker, K.~Monig, J.~Stelzer,
  Eur.\ Phys.\ J.\  {\bf C60}, 543-583 (2009).
  [arXiv:0811.0009 [hep-ph]].

\bibitem{Abazov:2011pt}
  V.~M.~Abazov {\it et al.} [ D0 Collaboration ],
  [arXiv:1104.2887 [hep-ex]].

\bibitem{Langenfeld:2009wd}
  U.~Langenfeld, S.~Moch, P.~Uwer,
  Phys.\ Rev.\  {\bf D80}, 054009 (2009).
  [arXiv:0906.5273 [hep-ph]].

\bibitem{Collins:1989gx}
  J.~C.~Collins, D.~E.~Soper, G.~F.~Sterman,
  Adv.\ Ser.\ Direct.\ High Energy Phys.\  {\bf 5}, 1-91 (1988).
  [hep-ph/0409313].

\bibitem{Nason:1987xz}
  P.~Nason, S.~Dawson, R.~K.~Ellis,
  Nucl.\ Phys.\  {\bf B303}, 607 (1988).

\bibitem{Beenakker:1988bq}
  W.~Beenakker, H.~Kuijf, W.~L.~van Neerven, J.~Smith,
  Phys.\ Rev.\  {\bf D40}, 54-82 (1989).

\bibitem{Beenakker:1990maa}
  W.~Beenakker, W.~L.~van Neerven, R.~Meng, G.~A.~Schuler, J.~Smith,
  Nucl.\ Phys.\  {\bf B351}, 507-560 (1991).

\bibitem{Czakon:2008ii}
  M.~Czakon, A.~Mitov,
  Nucl.\ Phys.\  {\bf B824}, 111-135 (2010).
  [arXiv:0811.4119 [hep-ph]].

\bibitem{Czakon:2007ej}
  M.~Czakon, A.~Mitov, S.~Moch,
  Phys.\ Lett.\  {\bf B651}, 147-159 (2007).
  [arXiv:0705.1975 [hep-ph]].

\bibitem{Czakon:2007wk}
  M.~Czakon, A.~Mitov, S.~Moch,
  Nucl.\ Phys.\  {\bf B798}, 210-250 (2008).
  [arXiv:0707.4139 [hep-ph]].

\bibitem{Czakon:2008zk}
  M.~Czakon,
  Phys.\ Lett.\  {\bf B664}, 307-314 (2008).
  [arXiv:0803.1400 [hep-ph]].

\bibitem{Bonciani:2008az}
  R.~Bonciani, A.~Ferroglia, T.~Gehrmann, D.~Maitre, C.~Studerus,
  JHEP {\bf 0807}, 129 (2008).
  [arXiv:0806.2301 [hep-ph]].

\bibitem{Bonciani:2009nb}
  R.~Bonciani, A.~Ferroglia, T.~Gehrmann, C.~Studerus,
  JHEP {\bf 0908}, 067 (2009).
  [arXiv:0906.3671 [hep-ph]].

\bibitem{Bonciani:2010mn}
  R.~Bonciani, A.~Ferroglia, T.~Gehrmann, A.~Manteuffel, C.~Studerus,
  JHEP {\bf 1101}, 102 (2011).
  [arXiv:1011.6661 [hep-ph]].

\bibitem{Korner:2008bn}
  J.~G.~Korner, Z.~Merebashvili, M.~Rogal,
  Phys.\ Rev.\  {\bf D77}, 094011 (2008).
  [arXiv:0802.0106 [hep-ph]].

\bibitem{Anastasiou:2008vd}
  C.~Anastasiou, S.~M.~Aybat,
  Phys.\ Rev.\  {\bf D78}, 114006 (2008).
  [arXiv:0809.1355 [hep-ph]].

\bibitem{Kniehl:2008fd}
  B.~Kniehl, Z.~Merebashvili, J.~G.~Korner, M.~Rogal,
  Phys.\ Rev.\  {\bf D78}, 094013 (2008).
  [arXiv:0809.3980 [hep-ph]].

\bibitem{Dittmaier:2007wz}
  S.~Dittmaier, P.~Uwer, S.~Weinzierl,
  Phys.\ Rev.\ Lett.\  {\bf 98}, 262002 (2007).
  [hep-ph/0703120 [HEP-PH]].

\bibitem{Dittmaier:2008yq}
  S.~Dittmaier, P.~Uwer, S.~Weinzierl,
  Nucl.\ Phys.\ Proc.\ Suppl.\  {\bf 183}, 196-201 (2008).
  [arXiv:0807.1223 [hep-ph]].

\bibitem{Bevilacqua:2010ve}
  G.~Bevilacqua, M.~Czakon, C.~G.~Papadopoulos, M.~Worek,
  Phys.\ Rev.\ Lett.\  {\bf 104}, 162002 (2010).
  [arXiv:1002.4009 [hep-ph]].

\bibitem{Melnikov:2010iu}
  K.~Melnikov, M.~Schulze,
  Nucl.\ Phys.\  {\bf B840}, 129-159 (2010).
  [arXiv:1004.3284 [hep-ph]].

\bibitem{Czakon:2010td}
  M.~Czakon,
  Phys.\ Lett.\  {\bf B693}, 259-268 (2010).
  [arXiv:1005.0274 [hep-ph]].

\bibitem{Czakon:2011ve}
  M.~Czakon,
  [arXiv:1101.0642 [hep-ph]].

\bibitem{Anastasiou:2010pw}
  C.~Anastasiou, F.~Herzog, A.~Lazopoulos,
  JHEP {\bf 1103}, 038 (2011).
  [arXiv:1011.4867 [hep-ph]].

\bibitem{Abelof:2011jv}
  G.~Abelof, A.~Gehrmann-De Ridder,
  JHEP {\bf 1104}, 063 (2011).
  [arXiv:1102.2443 [hep-ph]].

\bibitem{Bernreuther:2011jt}
  W.~Bernreuther, C.~Bogner, O.~Dekkers,
  [arXiv:1105.0530 [hep-ph]].

\bibitem{Sterman:1986aj}
  G.~F.~Sterman,
  Nucl.\ Phys.\  {\bf B281}, 310 (1987).

\bibitem{Catani:1989ne}
  S.~Catani, L.~Trentadue,
  Nucl.\ Phys.\  {\bf B327}, 323 (1989).

\bibitem{Moch:2008qy}
  S.~Moch, P.~Uwer,
  Phys.\ Rev.\  {\bf D78}, 034003 (2008).
  [arXiv:0804.1476 [hep-ph]].

\bibitem{Czakon:2009zw}
  M.~Czakon, A.~Mitov, G.~F.~Sterman,
  Phys.\ Rev.\  {\bf D80}, 074017 (2009).
  [arXiv:0907.1790 [hep-ph]].

\bibitem{Beneke:2009ye}
  M.~Beneke, M.~Czakon, P.~Falgari, A.~Mitov, C.~Schwinn,
  Phys.\ Lett.\  {\bf B690}, 483-490 (2010).
  [arXiv:0911.5166 [hep-ph]].

\bibitem{Ahrens:2009uz}
  V.~Ahrens, A.~Ferroglia, M.~Neubert, B.~D.~Pecjak, L.~L.~Yang,
  Phys.\ Lett.\  {\bf B687}, 331-337 (2010).
  [arXiv:0912.3375 [hep-ph]].

\bibitem{Ahrens:2010zv}
  V.~Ahrens, A.~Ferroglia, M.~Neubert, B.~D.~Pecjak, L.~L.~Yang,
  JHEP {\bf 1009}, 097 (2010).
  [arXiv:1003.5827 [hep-ph]].

\bibitem{Kidonakis:2010dk}
  N.~Kidonakis,
  Phys.\ Rev.\  {\bf D82}, 114030 (2010).
  [arXiv:1009.4935 [hep-ph]].

\bibitem{Ahrens:2011mw}
  V.~Ahrens, A.~Ferroglia, M.~Neubert, B.~D.~Pecjak, L.~L.~Yang,
  [arXiv:1103.0550 [hep-ph]].

\bibitem{Hoang:2000yr}
  A.~H.~Hoang, M.~Beneke, K.~Melnikov, T.~Nagano, A.~Ota, A.~A.~Penin, A.~A.~Pivovarov, A.~Signer {\it et al.},
  Eur.\ Phys.\ J.\ direct {\bf C2}, 1 (2000).
  [hep-ph/0001286].

\bibitem{Hoang:1999zc}
  A.~H.~Hoang, T.~Teubner,
  Phys.\ Rev.\  {\bf D60}, 114027 (1999).
  [hep-ph/9904468].

\bibitem{Bauer:2000yr}
  C.~W.~Bauer, S.~Fleming, D.~Pirjol, I.~W.~Stewart,
  Phys.\ Rev.\  {\bf D63}, 114020 (2001).
  [hep-ph/0011336].

\bibitem{Bauer:2001yt}
  C.~W.~Bauer, D.~Pirjol, I.~W.~Stewart,
  Phys.\ Rev.\  {\bf D65}, 054022 (2002).
  [hep-ph/0109045].

\bibitem{Beneke:2002ph}
  M.~Beneke, A.~P.~Chapovsky, M.~Diehl, T.~Feldmann,
  Nucl.\ Phys.\  {\bf B643}, 431-476 (2002).
  [hep-ph/0206152].

\bibitem{Becher:2006nr}
  T.~Becher, M.~Neubert,
  Phys.\ Rev.\ Lett.\  {\bf 97}, 082001 (2006).
  [hep-ph/0605050].

\bibitem{Ferroglia:2009ii}
  A.~Ferroglia, M.~Neubert, B.~D.~Pecjak, L.~L.~Yang,
  JHEP {\bf 0911}, 062 (2009).
  [arXiv:0908.3676 [hep-ph]].

\bibitem{Martin:2009iq}
  A.~D.~Martin, W.~J.~Stirling, R.~S.~Thorne, G.~Watt,
  Eur.\ Phys.\ J.\  {\bf C63}, 189-285 (2009).
  [arXiv:0901.0002 [hep-ph]].

\bibitem{Nadolsky:2008zw}
  P.~M.~Nadolsky, H.~-L.~Lai, Q.~-H.~Cao, J.~Huston, J.~Pumplin, D.~Stump, W.~-K.~Tung, C.~-P.~Yuan,
  Phys.\ Rev.\  {\bf D78}, 013004 (2008).
  [arXiv:0802.0007 [hep-ph]].

\bibitem{Martin:2009bu}
  A.~D.~Martin, W.~J.~Stirling, R.~S.~Thorne, G.~Watt,
  Eur.\ Phys.\ J.\  {\bf C64}, 653-680 (2009).
  [arXiv:0905.3531 [hep-ph]].

\bibitem{Lai:2010nw}
  H.~-L.~Lai, J.~Huston, Z.~Li, P.~Nadolsky, J.~Pumplin, D.~Stump, C.~-P.~Yuan,
  Phys.\ Rev.\  {\bf D82}, 054021 (2010).
  [arXiv:1004.4624 [hep-ph]].

\bibitem{Bigi:1994em}
  I.~I.~Y.~Bigi, M.~A.~Shifman, N.~G.~Uraltsev, A.~I.~Vainshtein,
  Phys.\ Rev.\  {\bf D50}, 2234-2246 (1994).
  [hep-ph/9402360].

\bibitem{Beneke:1994sw}
  M.~Beneke, V.~M.~Braun,
  Nucl.\ Phys.\  {\bf B426}, 301-343 (1994).
  [hep-ph/9402364].

\bibitem{Chetyrkin:1999ys}
  K.~G.~Chetyrkin, M.~Steinhauser,
  Phys.\ Rev.\ Lett.\  {\bf 83}, 4001-4004 (1999).
  [hep-ph/9907509].

\bibitem{Aliev:2010zk}
  M.~Aliev, H.~Lacker, U.~Langenfeld, S.~Moch, P.~Uwer, M.~Wiedermann,
  Comput.\ Phys.\ Commun.\  {\bf 182}, 1034-1046 (2011).
  [arXiv:1007.1327 [hep-ph]].

\end{thebibliography}
\end{document}